# Ultrawide-band Unidirectional Surface Plasmon Polariton Launchers

by *Cuicui Lu*, *Xiaoyong Hu*,* *Hong Yang*, and *Qihuang Gong**

Dr. C. C. Lu, Prof. X. Y. Hu, Prof. H. Yang, Prof. Q. H. Gong

State Key Laboratory for Mesoscopic Physics,

Department of Physics,

Peking University, Beijing 100871 (PR China)

E-mail: xiaoyonghu@pku.edu.cn, qhgong@pku.edu.cn

**ABSTRACT:**

Plasmonic devices and circuits, bridging the gap between integrated photonic and microelectronic technology, are promising candidates to realize on-chip ultrawide-band and ultrahigh-speed information processing. Unfortunately, the wideband surface plasmon source, one of the most important core components of integrated plasmonic circuits, is still unavailable up to now. This has seriously restricted the practical applications of plasmonic circuits. Here, we report an ultrawide-band unidirectional surface plasmon polariton launcher with high launching efficiency ratio and large extinction ratio, realized by combining plasmonic bandgap engineering and linear interference effect. This device offers excellent performances over an ultrabroad wavelength range from 690 to 900 nm, together with a high average launching efficiency ratio of 1.25, large average extinction ratio of 30 dB, and ultracompact lateral dimension of less than 4 μm. Compared with previous reports, the operating bandwidth is enlarged 210 folds, while the largest launching efficiency ratio, largest extinction ratio, and small feature size are maintained simultaneously. This provides a strategy for constructing on-chip surface plasmon source, and also paving the way for the study of integrated plasmonic circuits.

Unidirectional surface plasmon polariton (SPP) launcher, performing functions of photon-to-SPP conversion and subsequent SPP launching in the required direction, acts as a kind of on-chip SPP source, having great potential applications in the fields of integrated plasmonic circuits and devices, such as broadband wavelength-division multiplexing devices, broadband routers and sorters, and even single-photon transistors.[1-3] It has three key characteristics: broad operating bandwidth, high SPP launching efficiency ratio in the desired direction, and large extinction ratio, which is defined as the ratio between the SPP intensity launched into the desired direction and that into the opposite direction. Two methods have been proposed to construct nanoscale unidirectional SPP launcher. The first approach is to adopt highly oblique incidence excitation of symmetrical subwavelength geometries, including nanoslits, nanogrooves, and nanoridges.[4,5] But this method has a stringent requirement of the oblique incidence angle, excitation position, and incident light wavelength, which greatly limits its practical applications.[6,7] The other approach is to use asymmetric configuration to help control the launching direction of SPPs based on linear interference effect.[8,9] However, only a single operating wavelength was achieved experimentally due to the difficulty in precisely controlling the optical phase difference within a broad frequency range.[10,11] For example, In 2007, Tejeira *et al.* reported an unidirectional SPP launching with a launching efficiency ratio of 2, operating at the wavelength of 800 nm.[12] Subsequently, Choi *et al.* achieved an unidirectional SPP launching with a extinction ratio of 10 at the wavelength of 800 nm.[13] Recently, Baron *et al.* obtained a unidirectional SPP launcher with a launching efficiency of 0.52 and an extinction ratio of 47, operating at the wavelength of 800 nm.[14] Chen *et al.* also reached unidirectional SPP launching with an extinction ratio of 30 at the operating wavelength of 830 nm.[15-17] Therefore, the obstacle of difficulty in simultaneously achieving a wideband, high launching efficiency ratio, and large extinction ratio for unidirectional SPP launcher has seriously restricted the study of ultrahigh speed and ultrawide-band information processing based on integrated plasmonic chips.[18]

Here, we experimentally demonstrate a novel nanoscale unidirectional SPP launcher with an ultrawide operating bandwidth, high launching efficiency ratio, and large extinction ratio simultaneously, realized by combining plasmonic bandgap engineering and linear interference effect. The nanoscale SPP launcher consists of a nanoslit connected with a one-dimensional chirped plasmonic crystal in its left-side etched in a gold film coated with an organic polymer polyvinyl alcohol (PVA) layer on silicon dioxide substrate, as shown in Fig. 1a. The nanoslit acts as a photon-to-SPP converter and part of the energy of the incident light is coupled into SPP modes.[19,20] The one-dimensional chirped plasmonic crystal is composed of arrays of nanogrooves with a monotonically increased lattice constant. Barnes *et al.* and Balci *et al.* have pointed out that a perfect one-dimensional plasmonic crystal provides a SPP stop band due to strong Bragg scattering effect when the lattice constant is nearly half the SPP wavelength.[21,22] Gradually increasing the lattice constant ensures that the edge of the stop band for the guided SPP modes varies with position along the one-dimensional chirped plasmonic crystal, which has been confirmed by Gan's measured results.[23] Accordingly, the frequency range of the SPP stop band of a one-dimensional chirped plasmonic crystal is much larger than that of the perfect one-dimensional plasmonic crystal, which has been confirmed by our previous simulations.[24] As a result, the one-dimensional chirped plasmonic crystal acts as a perfect Bragg reflection mirror with an ultrawide bandwidth, and the left-propagating SPPs are inhibited in an ultrabroad wavelength range. Moreover, there exist a linear interference between the reflected SPPs by the chirped plasmonic crystal and the original right-propagating SPPs. As for the engineered plasmonic bandgap, the cut off frequency (corresponding to the edge of the stop band) varies along the chirped plasmonic crystal. As a result, the optical phase difference is different for different SPP modes. This makes it possible to reach a nearly constructive interference within an ultra-broad wavelength range, which ensures a high launching efficiency ratio, and large extinction ratio. A nanoscale unidirectional SPP launcher was realized experimentally with an ultrawide operating bandwidth of 210 nm, a high average launching efficiency ratio of 1.25, and a large average intensity contrast ratio of 30 dB simultaneously. The lateral dimensional was less than 4 μm. This compact device nearly meets the requirement for practical on-chip applications, and also paves the way for the study of ultrawide-band ultrahigh and speed information processing based on integrated plasmonic circuits.

The gold film was fabricated by using a laser molecular beam epitaxy (LMBE) growth system (Model LMBE 450, SKY Company, China). The beam (with a wavelength of 248 nm and a pulse repetition rate of 5 Hz) output from an excimer laser system (Model COMPexPro 205, Coherent Company, USA) was used as the excitation light source. The beam was focused onto a gold target mounted on a rotating holder, 15 mm away from the silicon dioxide substrate. The typical energy density of the excitation laser was about 500 mJ/cm$^2$. The growth rate measured by a film thickness/rate monitor was about 0.01 nm/pulse. PVA powder with an average molecular weight of 30,000 (China) was dissolved in de-ionized water with a weight ratio of 1:32. The spin coating method was adopted to fabricate the PVA layer on the surface of gold films. A focused-ion-beam (FIB) etching system (Model Helios NanoLab 600, FEI

Company, USA) was employed to prepare the patterns of the nanoslit and the one-dimensional chirped plasmonic crystal. The scanning-electron-microscopy (SEM) image of the SPP launcher sample is shown in Fig. 1b. The thickness was 300 nm for gold film and 150 nm for PVA layer. The depth, width, and length of the nanoslit were 450 nm, 200 nm, and 20 μm, respectively. The depth, width, and length of nanogrooves were 250 nm, 140 nm and 10 μm, respectively. To engineer the plasmonic bandgap, the lattice constant was increased monotonically from 320 to 400 nm in 10-nm increments, which meets the adiabatic condition and ensures that the SPP stop bandedge changes with position along the one-dimensional chirped plasmonic crystal.[24] The patterned area has a width of 3.38 μm for the one-dimensional chirped plasmonic crystal. The length of the patterned area of the one-dimensional chirped plasmonic crystal was only half of the nanoslit length. The upper part of the nanoslit, i.e. the isolated nanoslit, was used as an on-chip reference. We also etched two identical decoupling grating in the left- and right-side of the SPP launcher to help couple the SPP modes to free space for the purpose of measurement. The distance of the nanoslit to its adjacent groove was 530 nm and to two decoupling gratings on both sides was 8 μm. To study the SPP propagation properties, we calculated SPP propagation length as a function of incident light wavelength for a 300-nm thick gold film coated with a 150-nm thick PVA layer by using the finite element method (adopting a commercial software package Comsol Multiphysics),[25] and the calculated results are shown in Fig. 1c. The wavelength-dependent complex dielectric function of gold was obtained from Ref. [26]. The refractive index of PVA was 1.5 in the visible and near infrared range.[17] When the incident light wavelength was larger than 780 nm, the propagation length of the generated SPP mode was larger than 7 μm. This confirms the perfect propagation properties of SPPs. In our experiment, a micro-spectroscopy measurement system was adopted to measure the functions of the SPP launcher. The nanoslit was normally illuminated from the back side using a p-polarized continuous wave (CW) Ti:sapphire laser beam with different wavelengths. The optical-thick gold film can prohibit the direct transmission of the incident laser beam. The incident laser beam was focused into a spot with a radius of about 50 µm, ensuring uniform illumination of the entire nanoslit. The line width of the laser spectrum curve was about 1.5 nm, which ensures that only needed quasi-monochromatic SPP mode can be excited by the SPP launcher.[25] The SPP mode was scattered by two decoupling gratings in the output ports. The scattered light was collected by a long working distance objective (Mitutoyo 20, NA=0.58) and then imaged onto a charge coupled device (CCD). To study the role of the 150-nm-thick PVA layer, we calculated the magnetic-field distribution in the plane of 75 nm above the gold film for an 800 nm incident light in Au/air (300-nm-thick gold film in air) and Au/PVA/air (300-nm-thick gold film coated with a 150-nm-thick PVA layer in air) structures by using the finite element method, and the calculated results are shown in Fig. 1d. There is small magnetic-field distribution of the SPP mode in the plane of 75 nm above the gold film for the Au/air structure, which indicates that the magnetic-field distribution of the SPP mode is mainly confined in the gold surface. The case is different for the Au/PVA/air structure, i.e. the structure used in our experiment. There is strong magnetic-field distribution of the SPP mode in the PVA layer, which implies that the magnetic-field distribution of the SPP mode is elevated from the gold film surface into the upper PVA layer. Therefore, compared with the Au/air structure, much stronger modulation of the SPP propagation can be reached when a plasmonic crystal is etched in the Au/PVA/air structure. This has been confirmed by Chen's measured results.[17]

To study the SPP launching properties, we measured CCD images of the SPP launcher sample under excitation of different CW lasers, and the measured results are shown in the leftmost column of Fig. 2 (detailed in Fig. S1 of the Supporting Information). The upper parts of two decoupling gratings correspond to the case of excitation of the isolated nanoslit. For the case of excitation of the isolated nanoslit, strong scattered light is observed from the upper parts of two decoupling gratings in both the left- and right-sides when the incident light wavelength changes from 690 to 900 nm. SPPs are generated equally on both sides of the nanoslit. Thus there simultaneously exist the left- and right-propagating SPPs due to the symmetric configuration of the isolated nanoslit. The lower parts of two decoupling gratings correspond to the case of excitation of the nanoslit connected with the one-dimensional chirped plasmonic crystal in its left-side. The case is different when the nanoslit connected with chirped plasmonic crystal is excited. Strong scattered light is observed only from the lower part of the decoupling grating in the right-side when the incident light wavelength changes from 690 to 900 nm. No scattered light is seen from the lower part of the decoupling grating in the left-side. This implies that there only exist right-propagating SPPs, and the left-propagating SPPs are completely suppressed. The unidirectional SPP launching is reached in a very wide wavelength range from 690 to 900 nm. The achieved operating bandwidth, 210 nm, is enlarged 210 folds compared with some previously reported results.[13,15-17] A perfect one-dimensional plasmonic crystal could provide a SPP stop band.[27] While for a one-dimensional chirped plasmonic crystal having a slowly increasing lattice constant, its SPP stop band could be tuned gradually with position along the plasmonic crystal.[23] This implies that for a given lattice constant, a particular stop band edge is engaged. A larger lattice constant provides a longer cut off wavelength (corresponding to the stop band edge).[24] So, the wavelength range of the stop band of the one-dimensional chirped plasmonic crystal is generally much wider than that of the perfect one-dimensional plasmonic crystal. The one-dimensional chirped plasmonic crystal serves as an ultrawide-band Bragg reflection mirror for SPPs. As a result, the left-propagating SPPs are reflected back completely, and subsequently propagate toward the decoupling grating in the right-side. The left-propagating SPPs are inhibited greatly. The measured intensity distributions of the scattered light from the decoupling grating, obtained from CCD images, are plotted in the middle column (for the left-side decoupling grating) and the rightmost column (for the right-side decoupling grating) of Fig. 2. For the upper parts of two decoupling gratings, the scattered signal intensities are nearly the same because the isolated nanoslit is a symmetric nanostructure. While for the lower parts of two decoupling gratings, the scattered signal intensities from the left-side decoupling grating is completely different from that of the right-side decoupling grating because of the asymmetric configuration of the SPP launcher. The scattered signal intensities from the left-side decoupling grating nearly decrease to zero. While the scattered signal intensities from the right-side decoupling grating remain strong. To further confirm the SPP launching properties, we also calculated the steady-state power flow distributions in the plane 10 nm above the upper surface of the gold film as a function of incident light wavelength by using the finite element method, and the calculated results are shown in Fig. 3 (detailed in Fig. S1 of the Supporting Information). The steady-state power flow distributions decrease exponentially both for the isolated nanoslit and for the nanoslit connected with the one-dimensional chirped plasmonic crystal. It is very clear that for the case of excitation of the isolated nanoslit, the steady-state power flow distribution is symmetric. There exist left- and right-going SPP modes simultaneously. While for the case of excitation of the nanoslit connected with the one-dimensional chirped plasmonic crystal, there are only right-propagating SPP modes and the power flow is zero for the left-propagating SPP modes in a wide excitation wavelength range from 690 to 900 nm. These evidences confirm that the SPP launcher has the unidirectional SPP launching function within an ultrawide bandwidth.

To further characterize the unidirectional launching properties, we also measured the extinction ratio between the intensity of right- and left-propagating SPPs as a function of incident laser wavelength, and the measured results are shown in Fig. 4a. The extinction ratio can be obtained from $10\log(I_R/I_L)$, where $I_R$ and $I_L$ are the scattered light intensity from the right- and left-side decoupling gratings, respectively, which are extracted from the measured CCD images of Fig. 2. The measured signal curve took on a flat-top line shape. The value of the extinction ratio maintained at 35 dB when the incident laser wavelength changed from 730 to 840 nm. The value of the extinction ratio decreased to 24 dB for the 690 and 900-nm incident laser. The reasons may lie in the small nanogroove number of the one-dimensional chirped plasmonic crystal. The left-propagating SPPs were reflected back completely by the one-dimensional chirped plasmonic crystal. The phase difference $\phi$ between the reflected SPP mode and original right-propagating SPP mode can be calculated by[12]

$$\phi = 2k_{\text{SPP}}d + m\pi \tag{1}$$

where $k_{\text{SPP}}$ is the SPP wavevector, $d$ is the distance between the center of the nanoslit and the center of the nanogrooves in the one-dimensional chirped plasmonic crystal where the cutoff wavelength equal to the SPP wavelength. A constructive (or destructive) interference should occur for values of the phase difference equal to even (or odd) multiples of π. The average extinction ratio reached 30 dB when the wavelength of the incident laser changed from 690 to 900 nm. This value was among the highest ones compared with previous experimental reports.[12-17] The perfect ultrawide-band Bragg refection provided by the chirped plasmonic crystal as well as the linear interference between the reflected SPPs and the original right-propagating SPPs guarantee a large average extinction ratio. The measured extinction ratios were in qualitative agreement with the calculated ones by using the finite element method, as shown in Fig. 4b. The difference between the measured and calculated results originates from large SPP propagation losses of the fabricated sample, caused by imperfectly etched grating nanostructure, as shown in Fig. 1b, and large surface roughness of 10 nm of the gold film, as shown in Fig. 1e. We also measured the launching efficiency ratio as a function of incident laser wavelength, and the measured results are shown in Fig. 4c. The launching efficiency ratio is defined as the quotient between the intensities of the right-propagating SPPs with and without the one-dimensional chirped plasmonic crystal.[12] In our experiment, the launching efficiency ratio was obtained by the quotient between the scattered intensities of right-side decoupling gratings with and without the one-dimensional chirped plasmonic crystal, which were extracted from the measured CCD images of Fig. 2. The measured signal curve took on an oscillatory line shape, which originates from the interference effect. The average SPP launching efficiency ratio was measured to be 1.25, which was also among the highest values compared with previous experimental reports.[12-17] This can also be confirmed from the calculated steady-state power flow distributions in the plane 10 nm above the upper surface of the gold film under excitation of different incident light, as shown in Fig. 3. The steady-state power flow distributions of the SPP launcher were much stronger than that of the isolated nanoslit under excitation of the same incident light. Therefore, a large average launching efficiency ratio was achieved. The measured results were in qualitative agreement with the calculated ones by using the finite element method, as shown in Fig. 4d. The difference between the measured and calculated results originates from the discrepancy in the structural parameters between the practical sample and the calculation model.

In conclusion, we realized a nanoscale unidirectional SPP launcher based on a nanoslit connected with a one-dimensional chirped plasmonic crystal. An ultrabroad operating bandwidth of 210 nm, a high average launching efficiency ratio of 1.25, and a large average extinction ratio of 30 dB were obtained simultaneously. The lateral dimensional was less than 4 μm. This compact device is robust, free from environment impact, and much suitable for practical on-chip applications. This may offer a strategy for constructing integrated plasmonic circuits and realizing ultrawide-band and ultrahigh speed information processing based on plasmonic chips.

**Acknowledgements**

This work was supported by the National Basic Research Program of China under grant 2013CB328704, the National Natural Science Foundation of China under grants 11225417, 61077027, 11134001, 11121091, and 90921008, and the program for New Century Excellent Talents in University.

**Figure Captions**

**Figure 1.** Characteristics of the SPP launcher. (a) Schematic cross-section structure. (b) Top-view SEM image. (c) Calculated SPP propagation length as a function of incident light wavelength for a 300-nm thick gold film coated with a 150-nm thick PVA layer. (d) Calculated magnetic field distribution in the plane of 75 nm above the gold film for an 800 nm incident light in the Au/air and Au/PVA/air structures. (e) AFM image of a 300-nm-thick gold film.

**Figure 2.** Measured CCD image of the SPP launcher (Left), measured intensity distribution of the scattered light from the decoupling grating in the left-side (Middle), and measured intensity distribution of the scattered light from the decoupling grating in the right-side (Right) under excitation of a CW laser with a wavelength of 690 nm (a), 750 nm (b), 800 nm (c), 850 nm (d), 900 nm (e).

**Figure 3.** Calculated steady-state power flow distribution in the plane 10 nm above the upper surface of the gold film as a function of incident light wavelength. (a) For 690 nm. (b) For 750 nm. (c) For 800 nm. (d) For 850 nm. (e) For 900 nm. Arrows indicate starting positions of the decoupling gratings in the left- and right-sides.

**Figure 4.** Characterization of the unidirectional SPP launching function. Measured (a) and calculated (b) extinction ratio as a function of incident laser wavelength. Measured (c) and calculated (d) launching efficiency ratio as a function of incident laser wavelength.

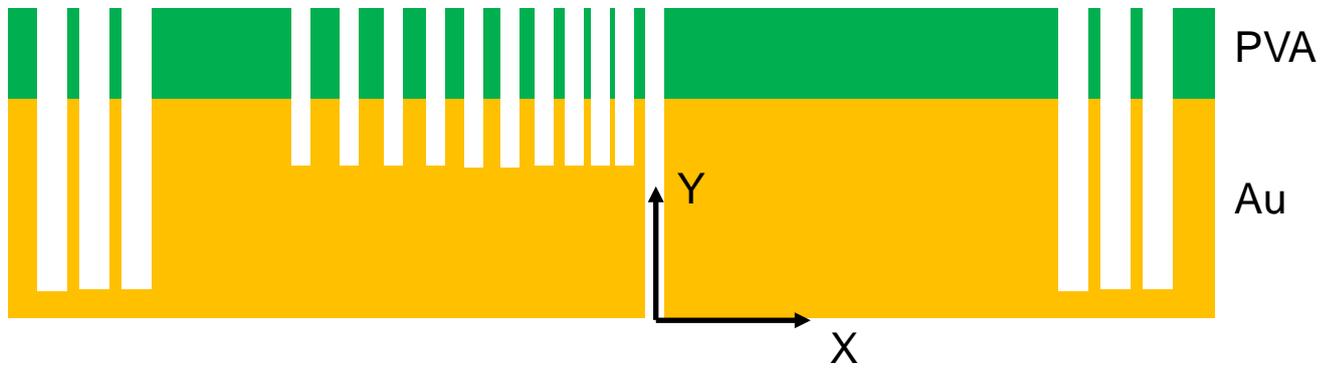

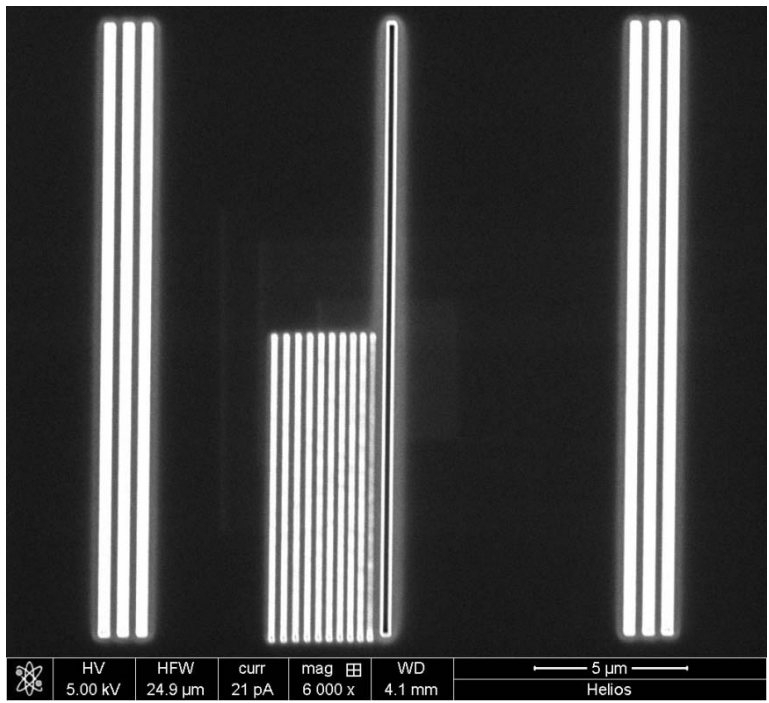

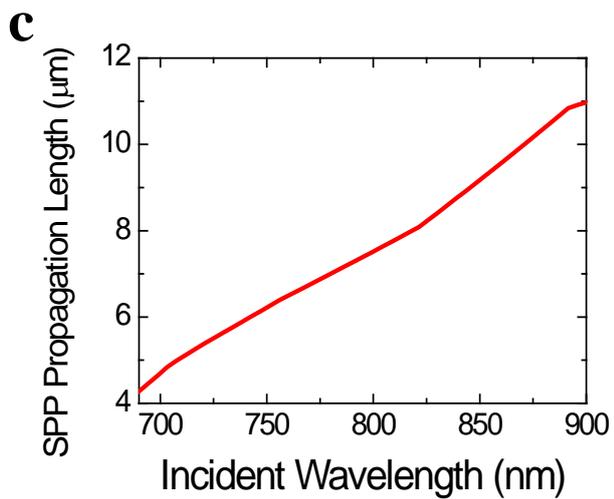
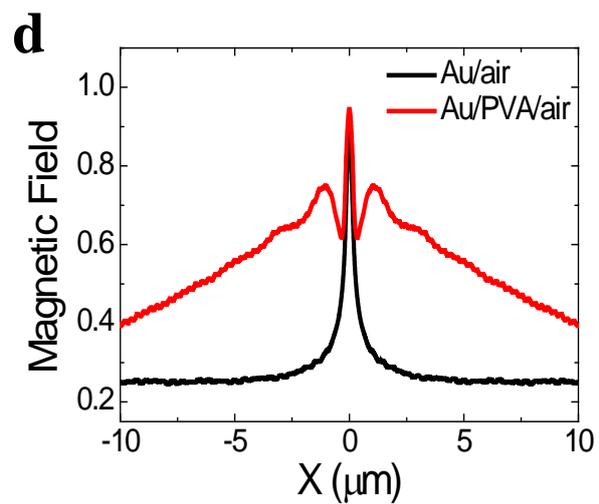
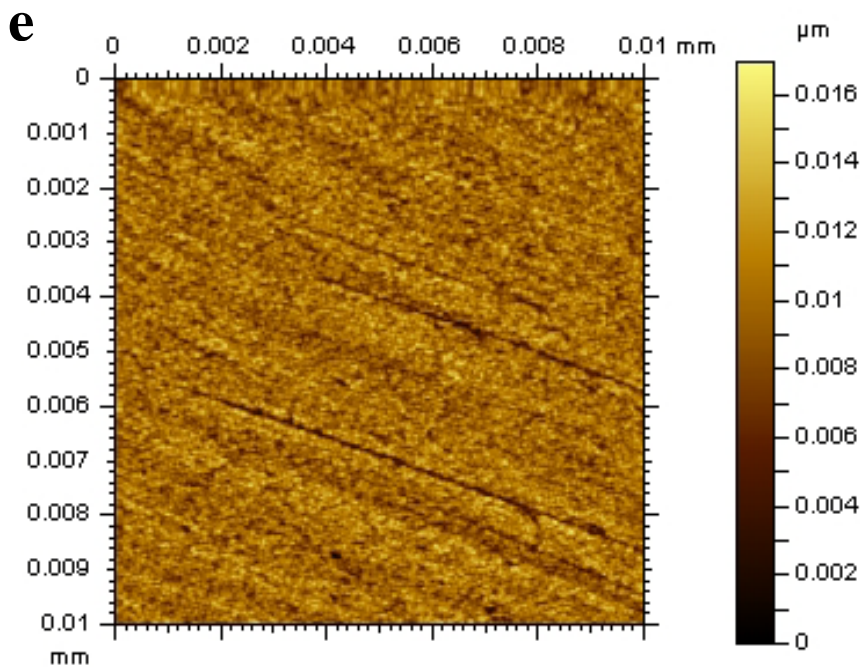

Figure 1. C. C. LU

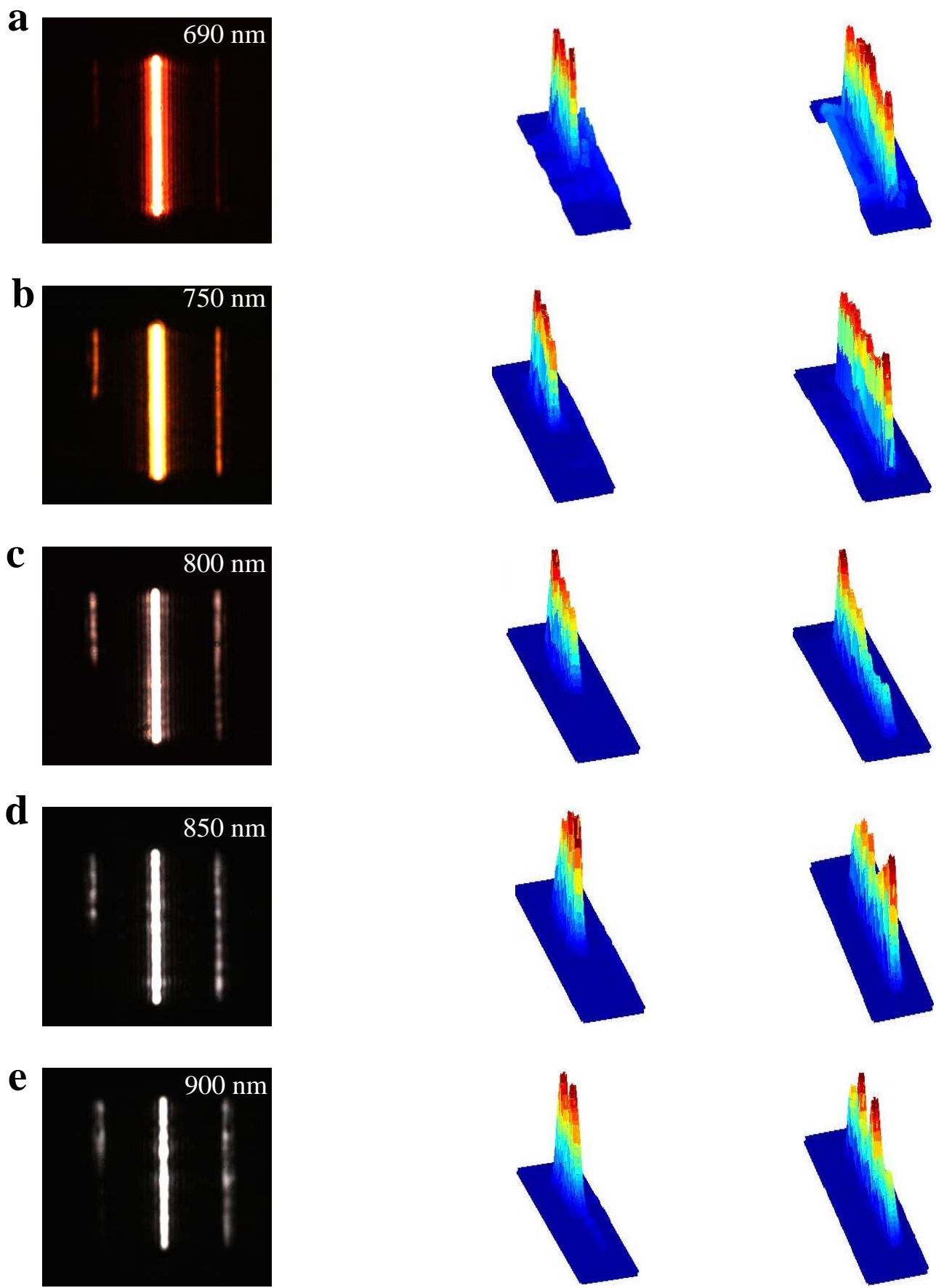

Figure 2. C. C. Lu

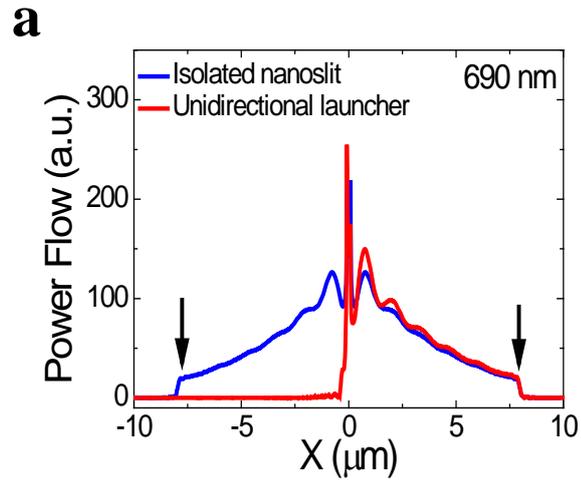
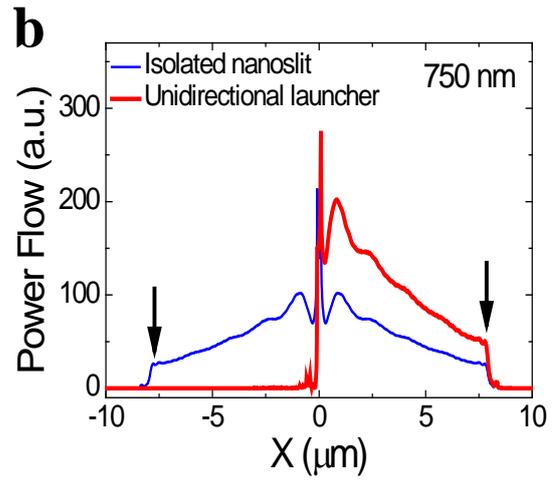
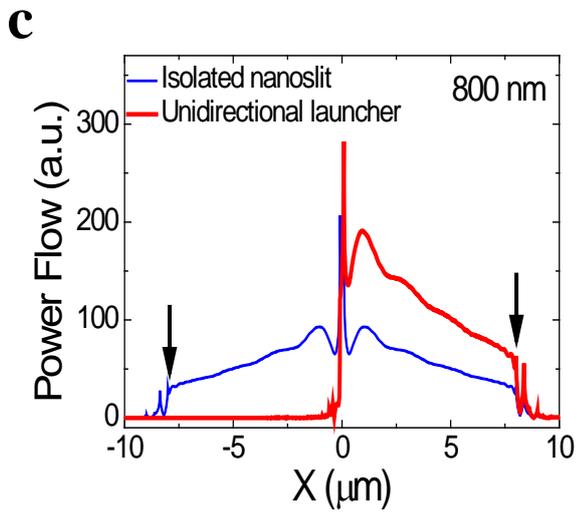
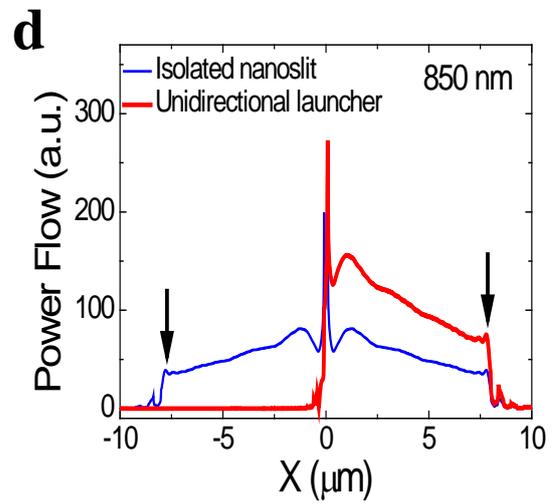
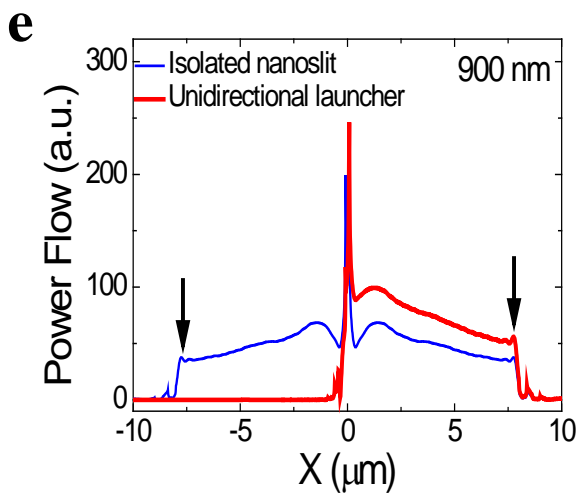

Figure 3. C. C. Lu

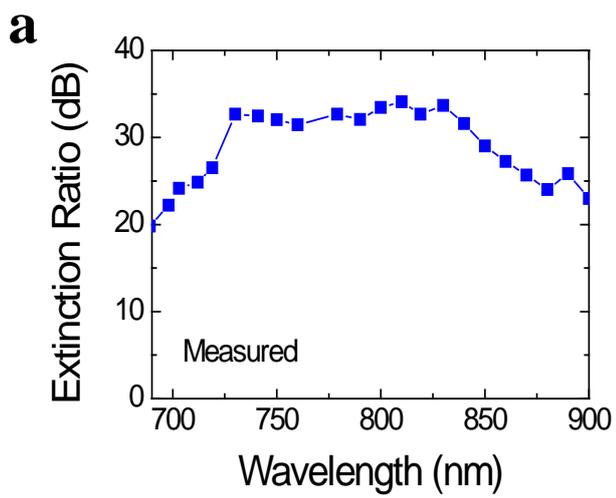
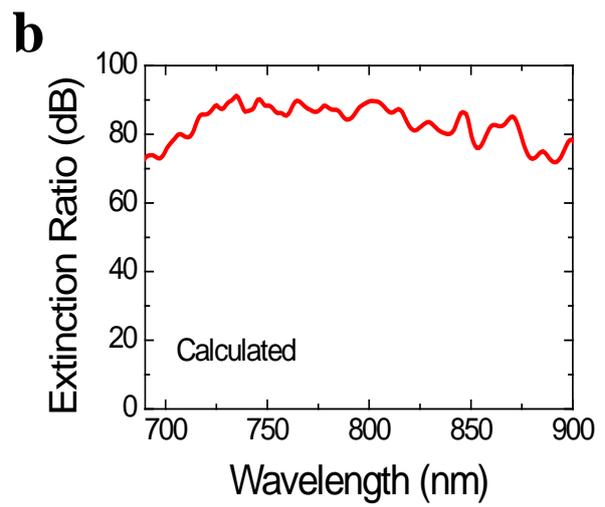
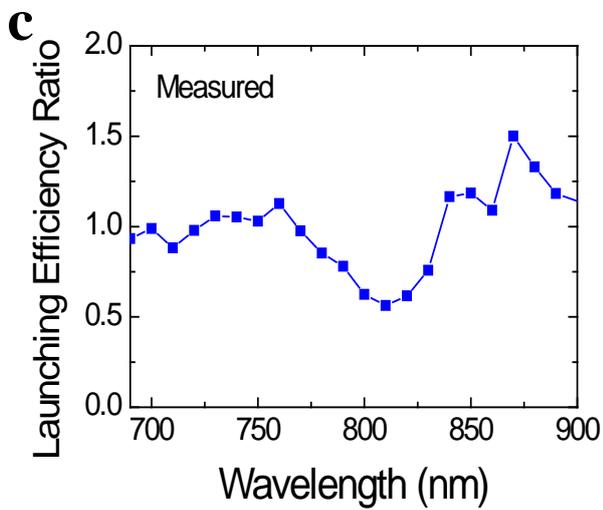
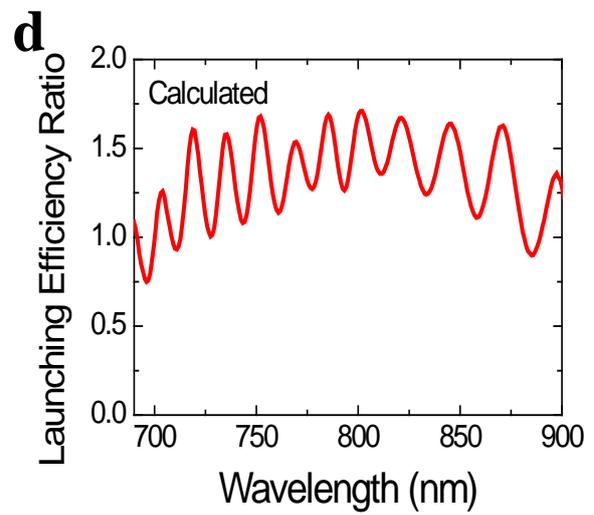

Figure 4. C. C. Lu

# Supporting Information

*More evidences confirming the ultrawide-band unidirectional SPP launching properties:*

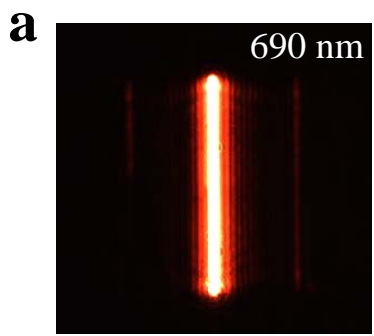 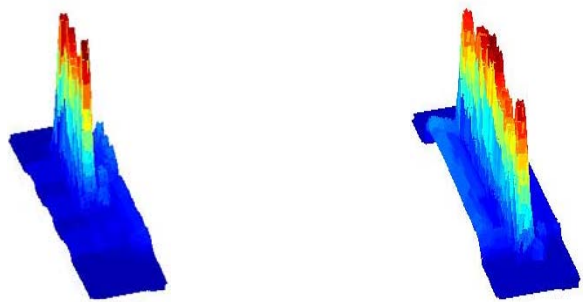

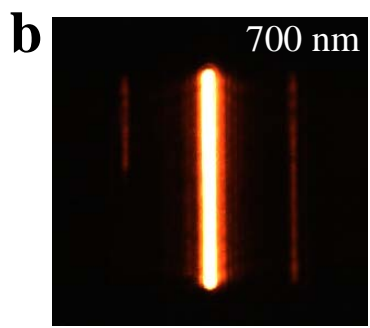 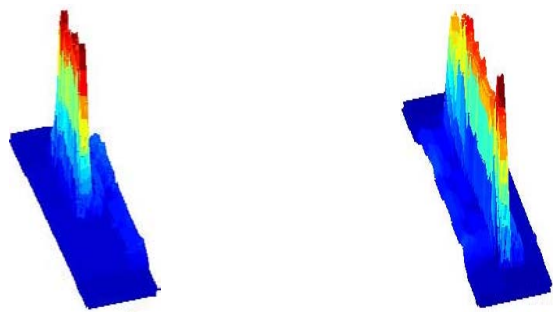

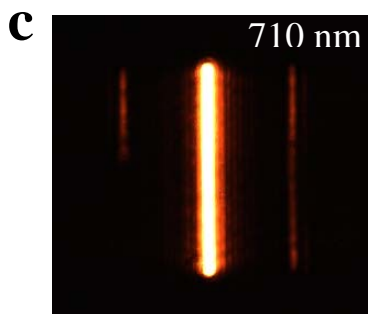 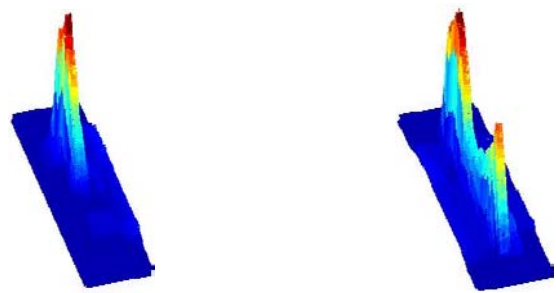

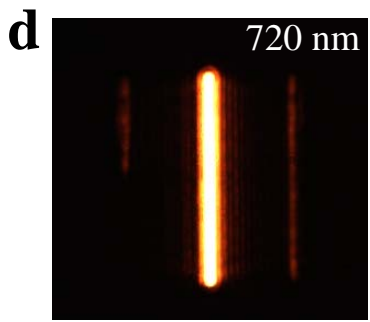 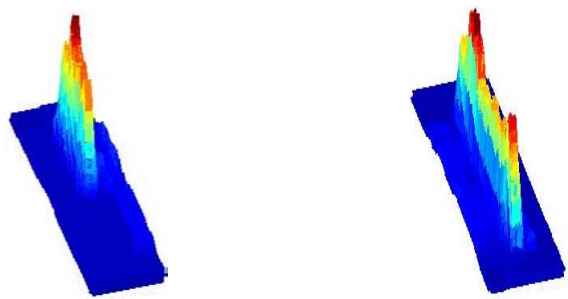

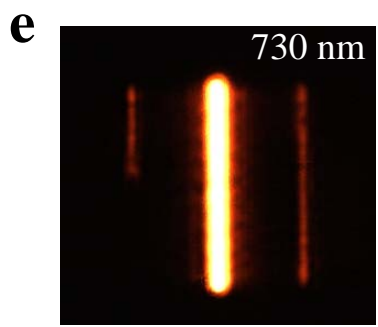
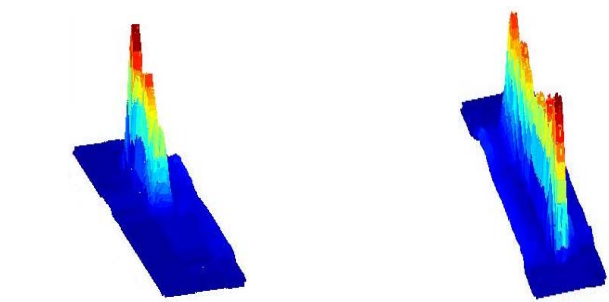

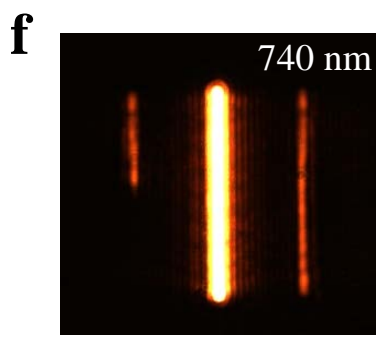
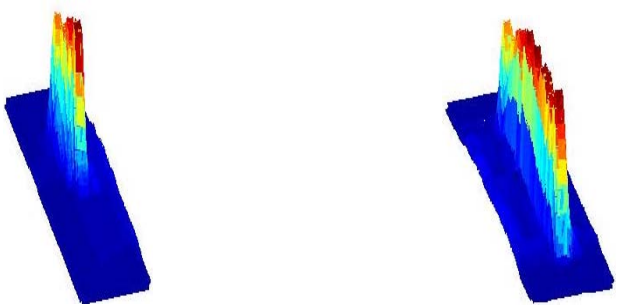

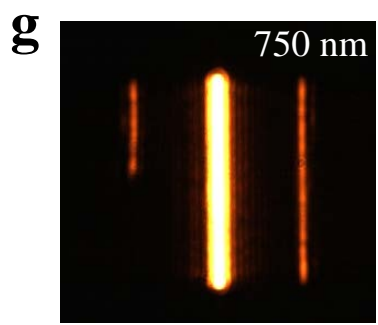
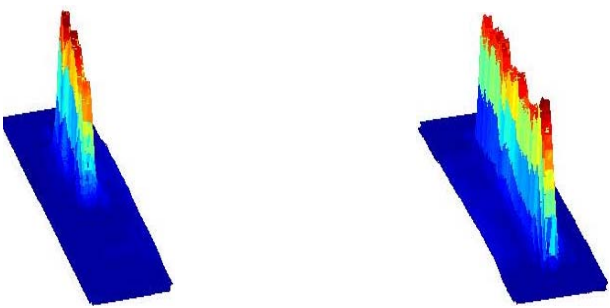

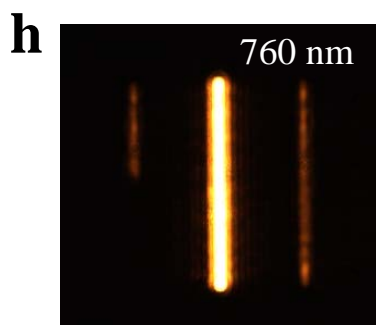
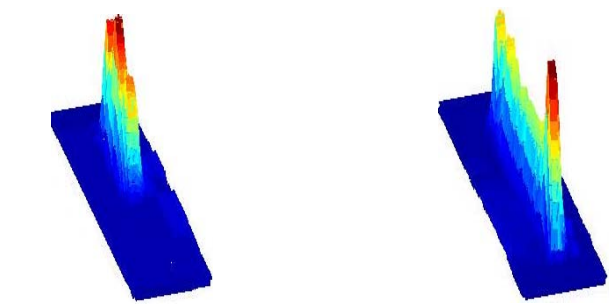

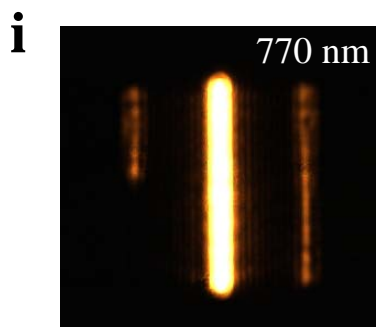
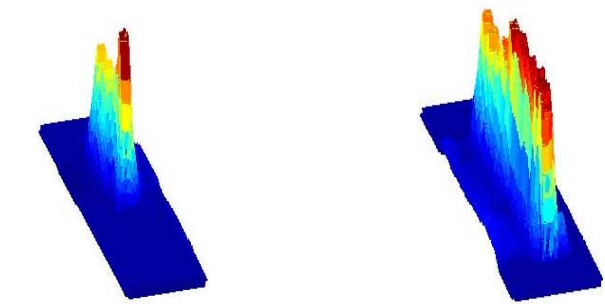

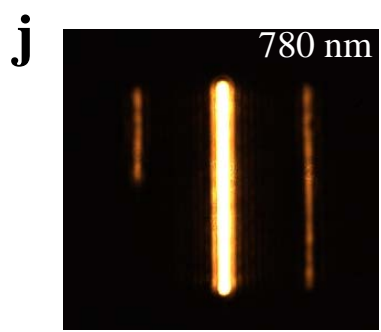 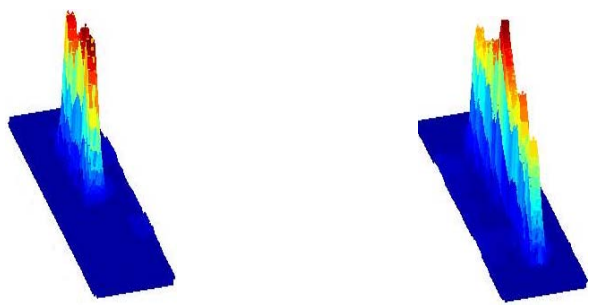

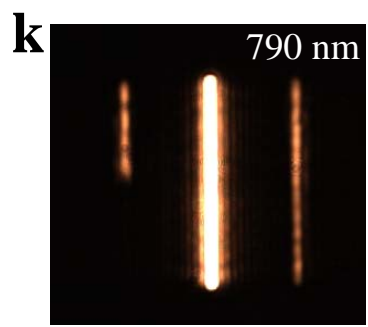 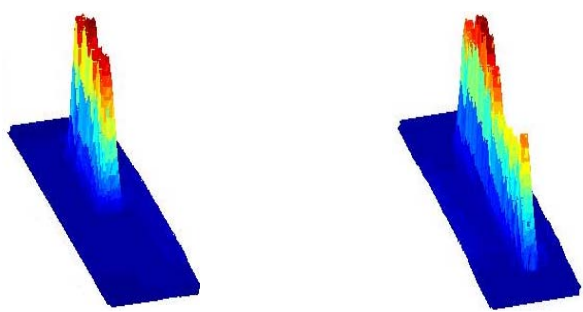

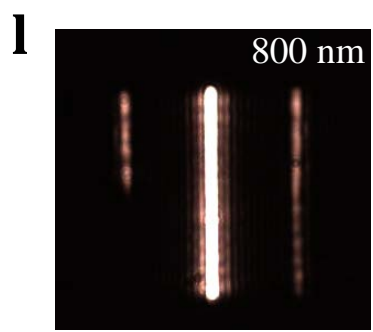 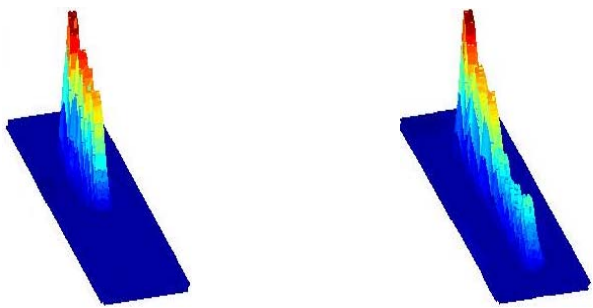

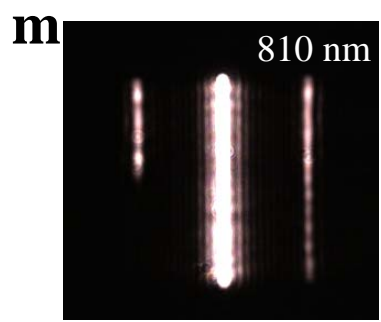 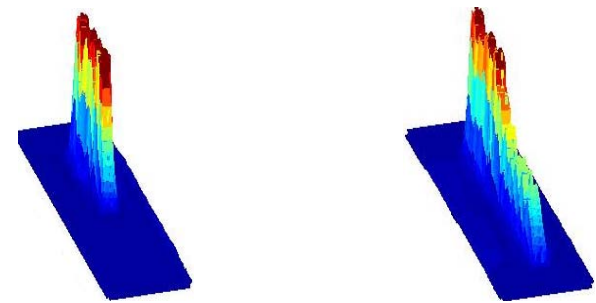

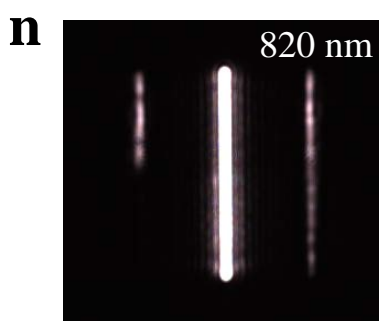 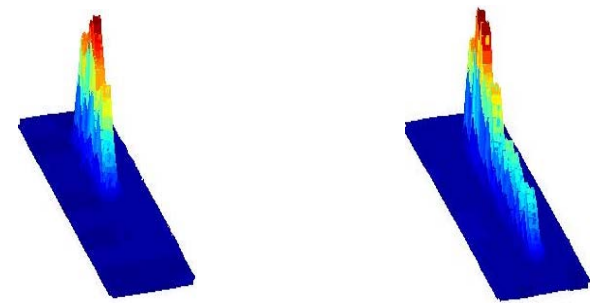

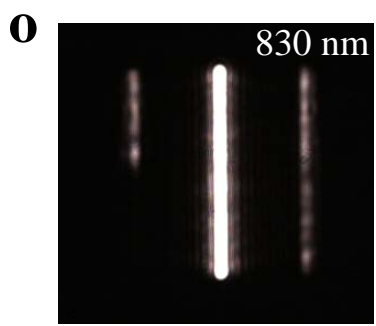
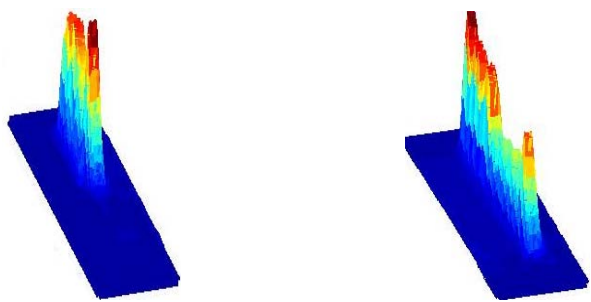

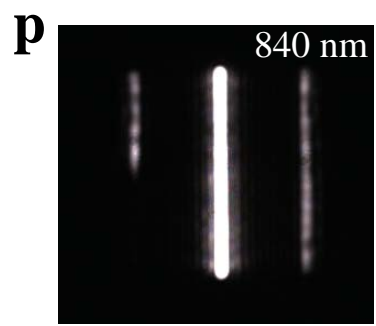
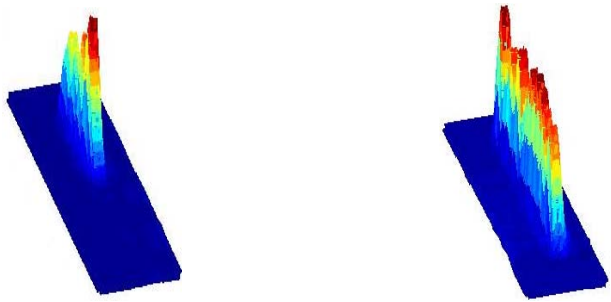

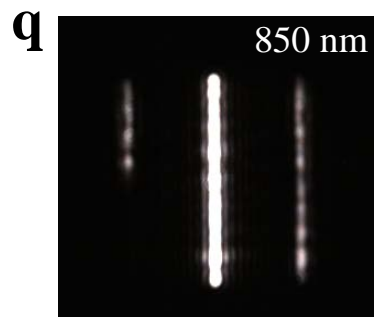
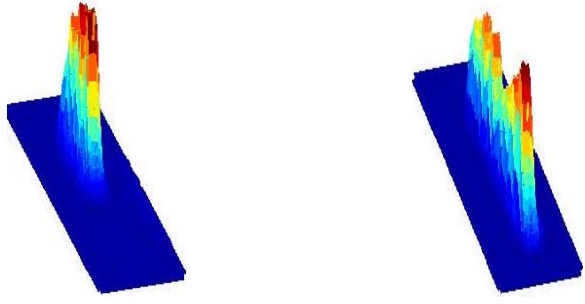

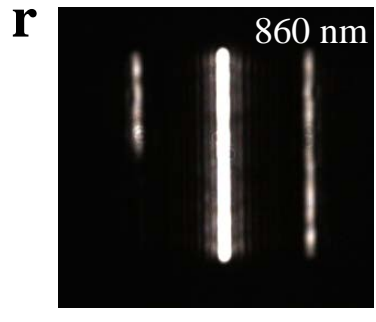
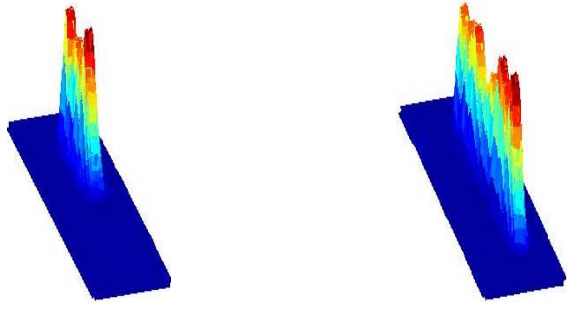

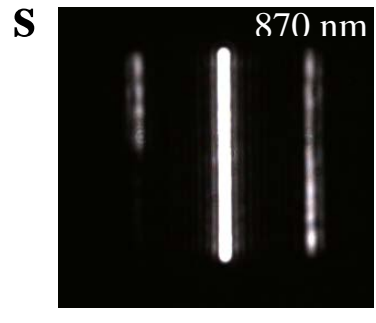
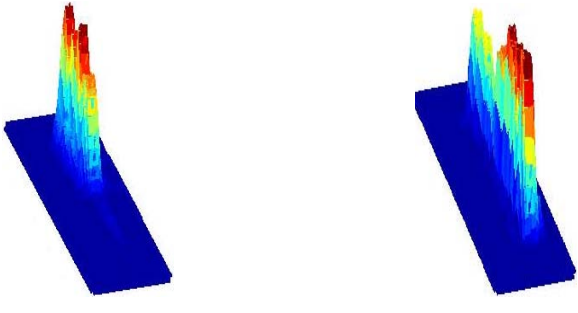

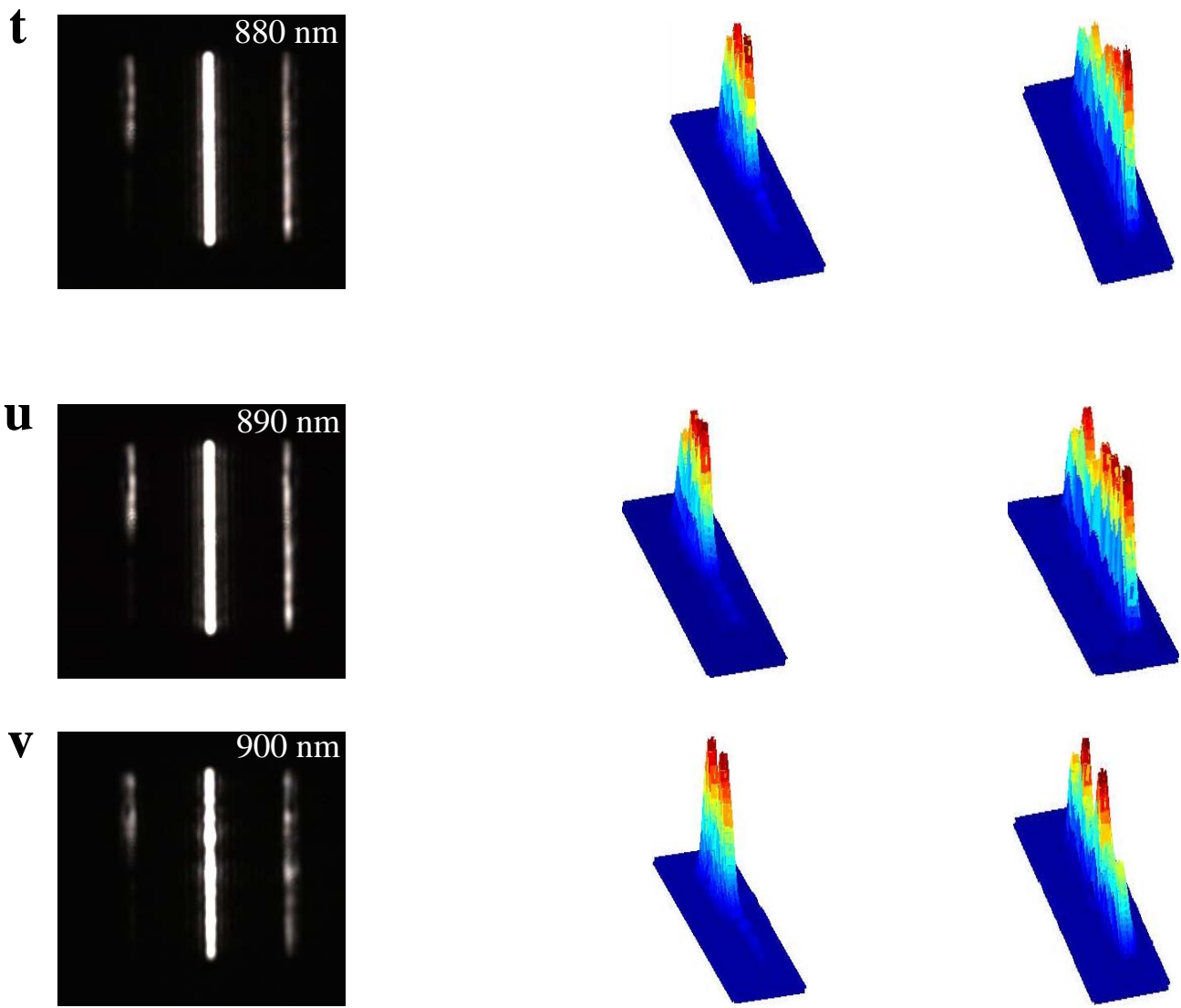

**Figure S1.** Measured CCD image of the SPP launcher (Left), measured intensity distribution of the scattered light from the decoupling grating in the left-side (Middle), and measured intensity distribution of the scattered light from the decoupling grating in the right-side (Right) under excitation of a CW laser with a wavelength of 690 nm (a), 700 nm (b), 710 nm (c), 720 nm (d), 730 nm (e), 740 nm (f), 750 nm (g), 760 nm (h), 770 nm (i), 780 nm (j), 790 nm (k), 800 nm (l), 810 nm (m), 820 nm (n), 830 nm (o), 840 nm (p), 850 nm (q), 860 nm (r), 870 nm (s), 880 nm (t), 890 nm (u), and 900 nm (v).

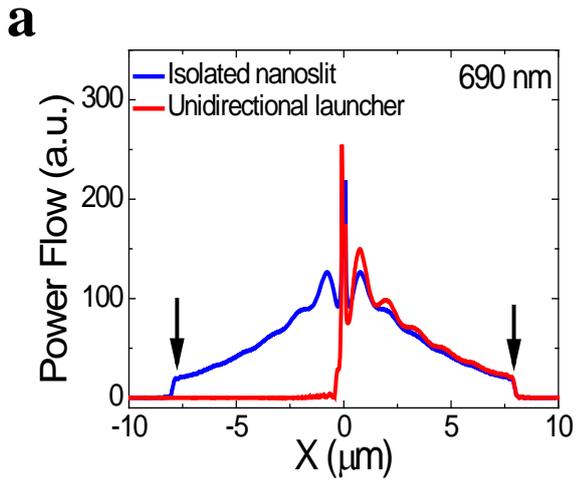
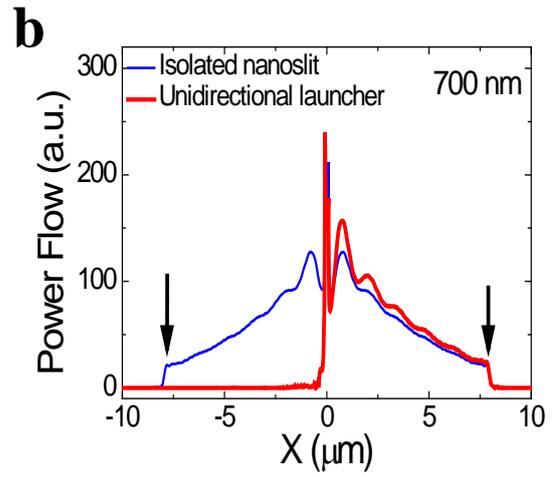
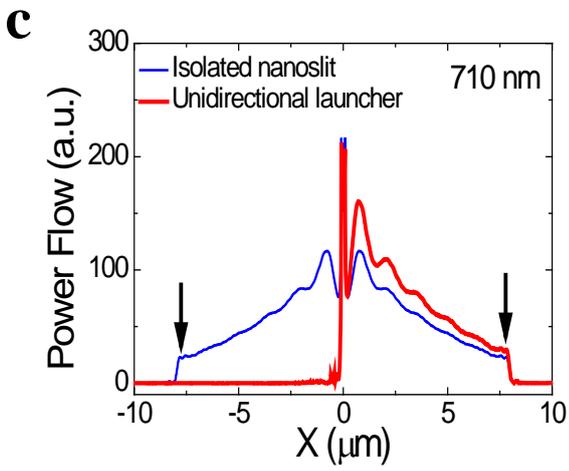
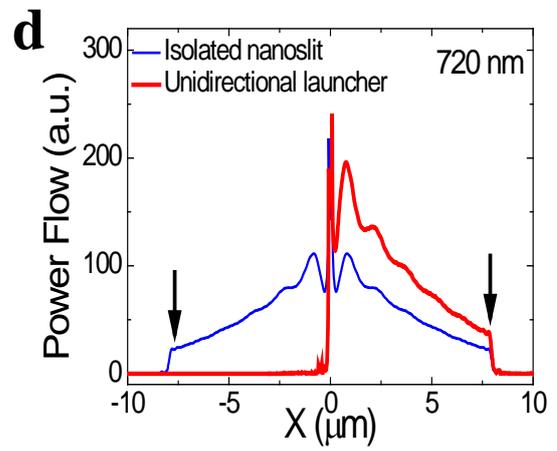
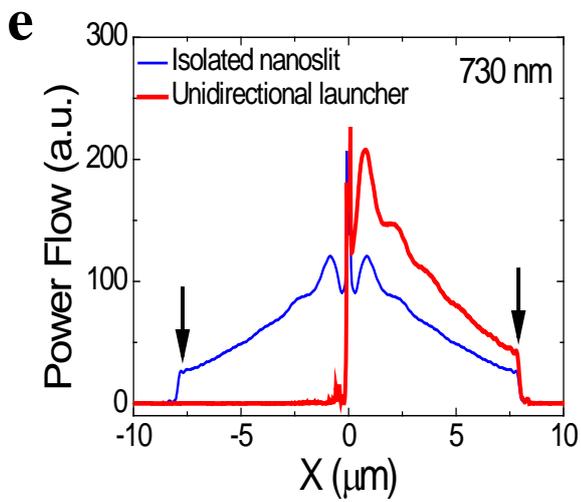
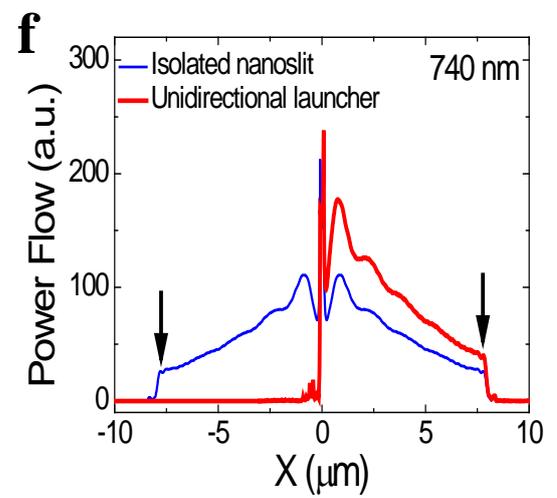

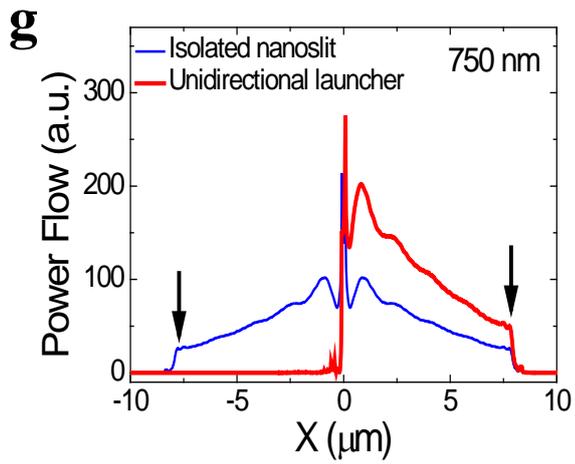
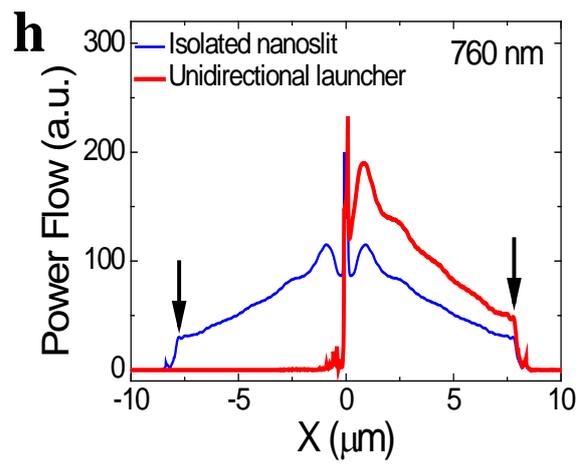
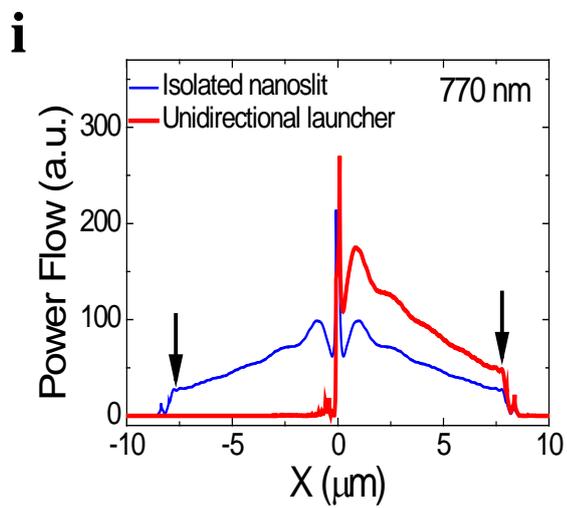
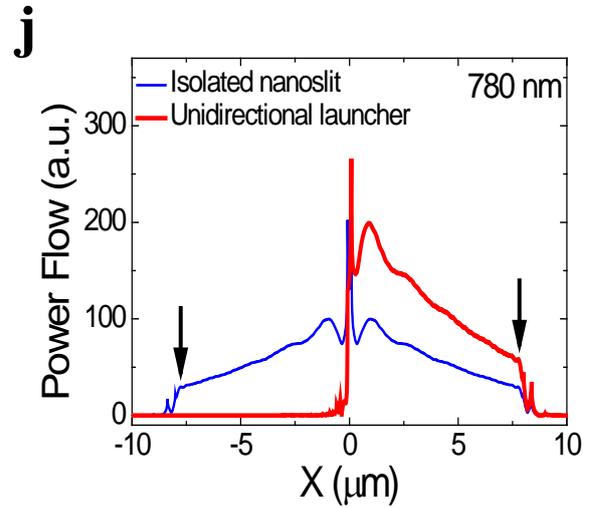
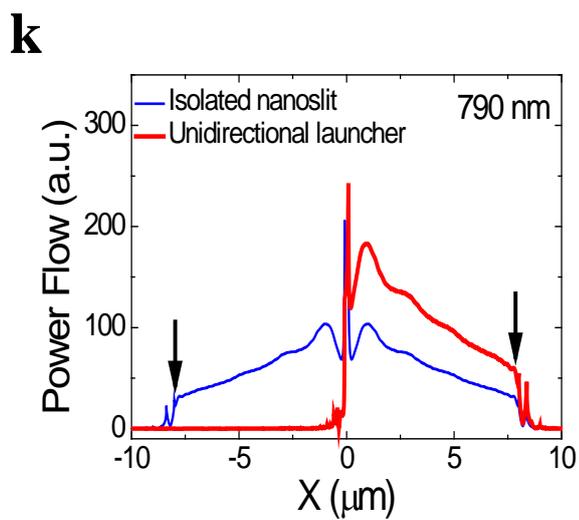
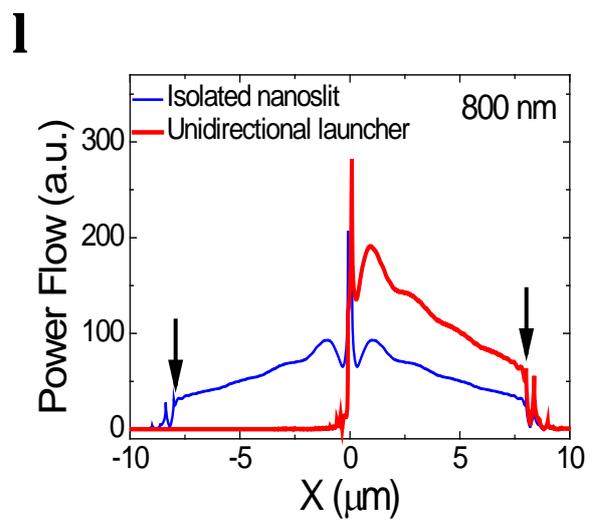

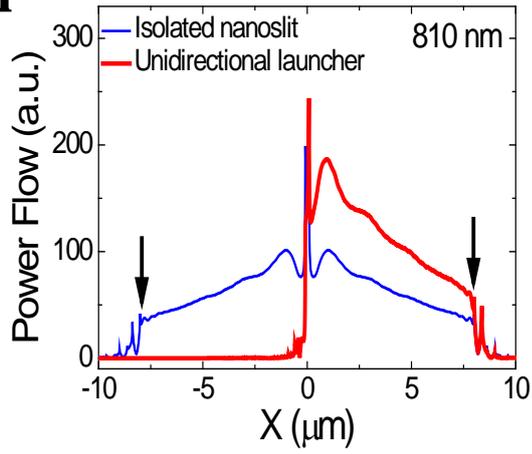
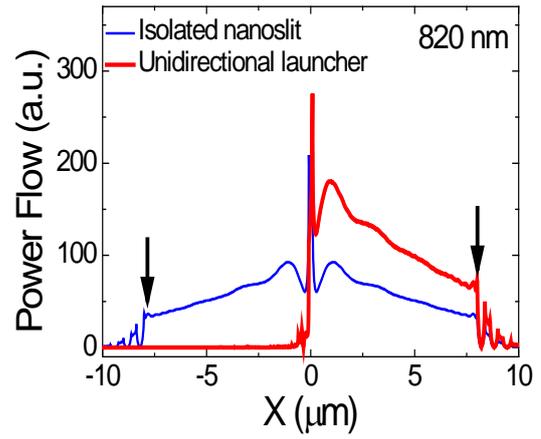
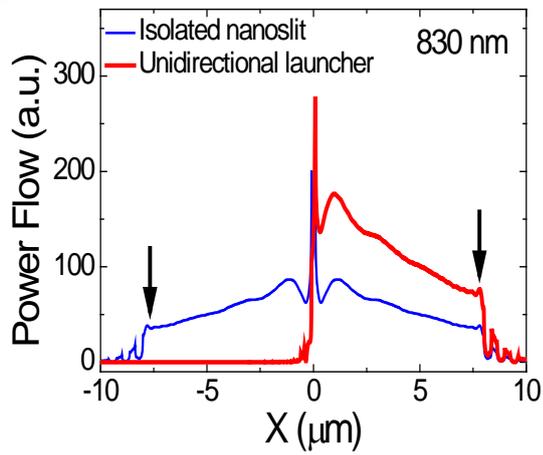
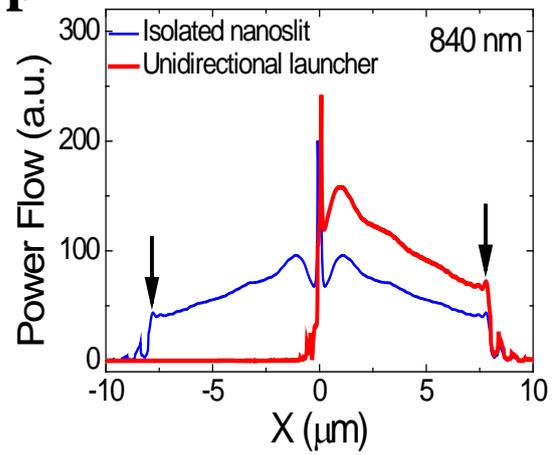
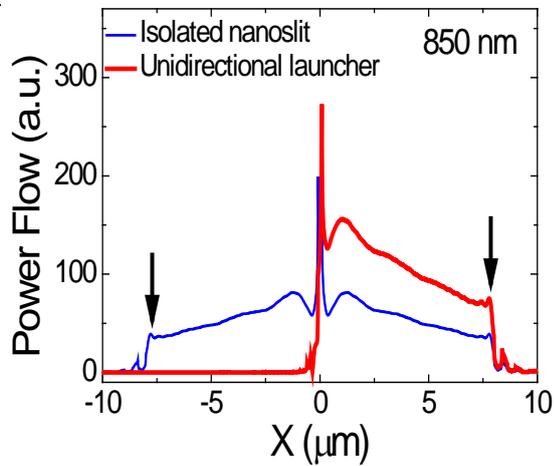
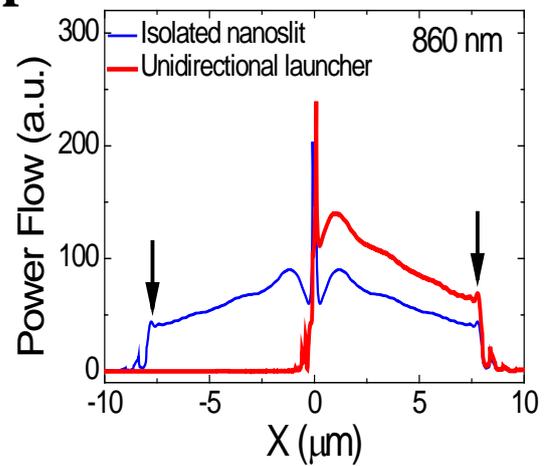

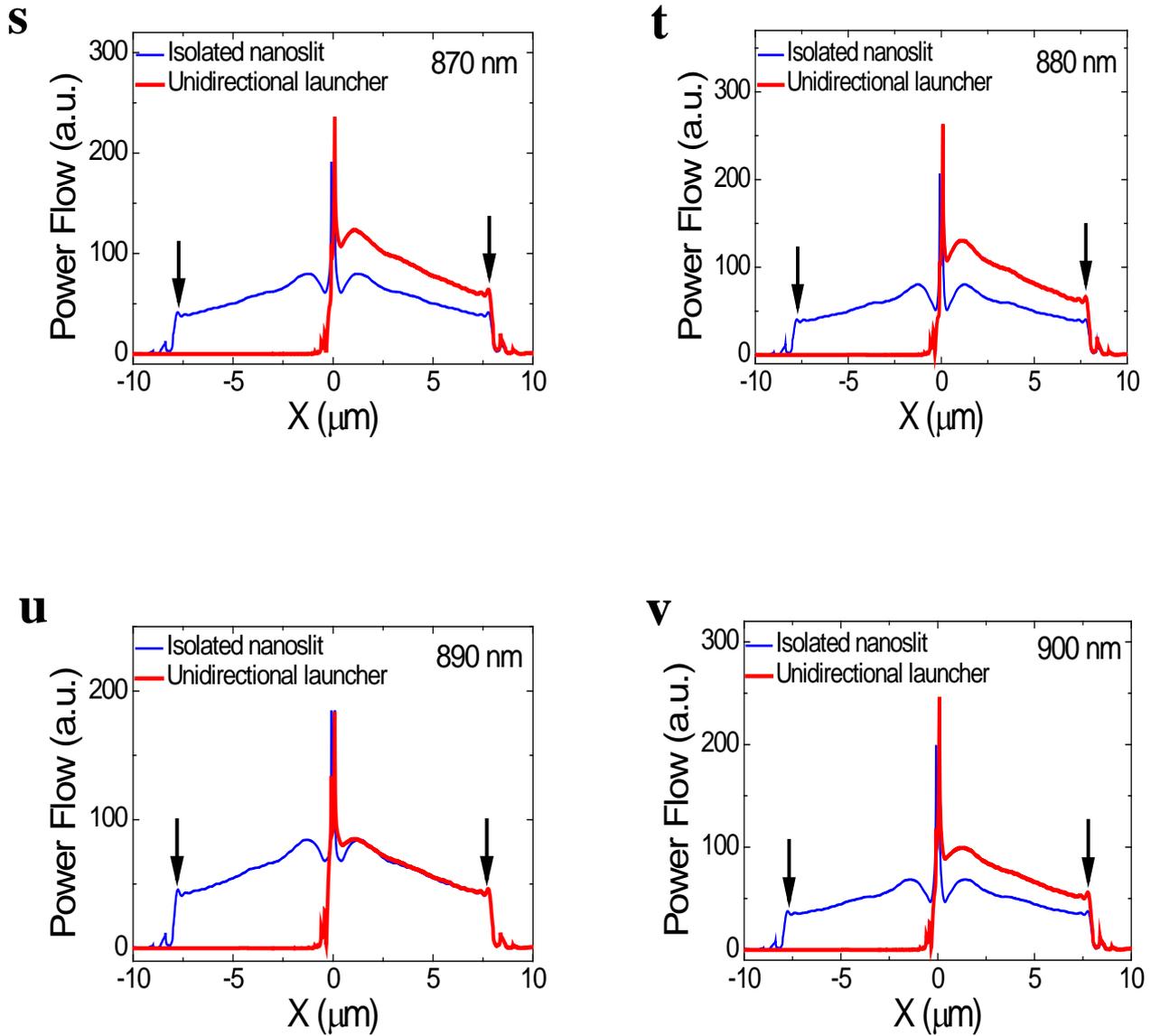

**Figure S2.** Calculated steady-state power flow distribution in the plane 10 nm above the upper surface of the gold film as a function of incident light wavelength. (a) For 690 nm. (b) For 700 nm. (c) For 710 nm. (d) For 720 nm. (e) For 730 nm. (f) For 740 nm. (g) For 750 nm. (h) For 760 nm. (i) For 770 nm. (j) For 780 nm. (k) For 790 nm. (l) For 800 nm. (m) 810 nm. (n) For 820 nm. (o) For 830 nm. (p) For 840 nm. (q) For 850 nm. (r) For 860 nm. (s) For 870 nm. (t) For 880 nm. (u) For 890 nm. (v) For 900 nm. Arrows indicate starting positions of the decoupling gratings in the left- and right-sides.